\documentclass[12pt, letterpaper]{extarticle}
\usepackage{hyperref}
\usepackage{url}
\usepackage{adjustbox}
\usepackage[flushleft]{threeparttable}
\usepackage{booktabs}
\usepackage{graphics}
\usepackage{helvet}
\usepackage{times}
\usepackage{comment}
\usepackage{placeins}
\usepackage{multirow}
\usepackage{rotating}
\usepackage{mathtools}
\usepackage{subfig}
\usepackage{arydshln}
\usepackage{float}
\usepackage{multicol}
\usepackage{supertabular}
\usepackage{booktabs}
\usepackage[usenames,dvipsnames,svgnames,table]{xcolor}
\usepackage[framemethod=TikZ]{mdframed}
\usepackage{fullpage}
\usepackage[switch]{lineno}
\usepackage{amsmath}
\usepackage{amssymb}
\usepackage{rotating}
\usepackage{array}
\usepackage{mathtools}
\usepackage[ruled]{algorithm2e}
\usepackage{algorithmic}
\usepackage{bm}
\usepackage{breqn}
\usepackage{comment}
\usepackage{enumitem}
\usepackage{graphics}
\usepackage{graphicx}
\usepackage{latexsym}
\usepackage{mathrsfs}
\usepackage{morefloats}
\usepackage{nicefrac}
\usepackage{authblk}
\usepackage{ulem}
\usepackage{cite}
\usepackage{xurl}

\definecolor{normgreen}{rgb}{0.0, 0.5, 0.0}

\title{Perception, performance, and detectability of conversational artificial intelligence across 32 university courses}

\author[1]{Hazem Ibrahim}
\author[1]{Fengyuan Liu}
\author[1]{Rohail Asim}
\author[1]{Balaraju Battu}
\author[1]{Sidahmed Benabderrahmane}
\author[1]{Bashar Alhafni}
\author[2]{Wifag Adnan}
\author[3]{Tuka Alhanai}
\author[2]{Bedoor AlShebli}
\author[1]{Riyadh Baghdadi}
\author[1]{Jocelyn J. Bélanger}
\author[1]{Elena Beretta}
\author[3]{Kemal Celik}
\author[1]{Moumena Chaqfeh}
\author[3]{Mohammed F. Daqaq}
\author[2]{Zaynab El Bernoussi}
\author[1]{Daryl Fougnie}
\author[3]{Borja Garcia de Soto}
\author[1]{Alberto Gandolfi}
\author[3]{Andras Gyorgy}
\author[1]{Nizar Habash}
\author[2]{J. Andrew Harris}
\author[2]{Aaron Kaufman}
\author[1]{Lefteris Kirousis}
\author[2]{Korhan Kocak}
\author[2]{Kangsan Lee}
\author[2]{Seungah S. Lee}
\author[2]{Samreen Malik}
\author[3]{Michail Maniatakos}
\author[1]{David Melcher}
\author[1]{Azzam Mourad}
\author[2]{Minsu Park}
\author[3]{Mahmoud Rasras}
\author[2]{Alicja Reuben}
\author[1]{Dania Zantout}
\author[2]{Nancy W. Gleason}
\author[2]{Kinga Makovi}
\author[1*]{Talal Rahwan}
\author[1*]{Yasir Zaki}

\affil[1]{\normalsize Division of Science, New York University Abu Dhabi, UAE}
\affil[2]{\normalsize Division of Social Science, New York University Abu Dhabi, UAE}
\affil[3]{\normalsize Division of Engineering, New York University Abu Dhabi, UAE}

\affil[*]{\footnotesize To whom correspondence should be addressed; E-mail: \{talal.rahwan, yasir.zaki\}@nyu.edu.}

\linenumbers

\date{}
\nolinenumbers
\begin{document}
\maketitle

\section*{Abstract}

The emergence of large language models has led to the development of powerful tools such as ChatGPT that can produce text indistinguishable from human-generated work. With the increasing accessibility of such technology, students across the globe may utilize it to help with their school work---a possibility that has sparked discussions on the integrity of student evaluations in the age of artificial intelligence (AI). To date, it is unclear how such tools perform compared to students on university-level courses. Further, students' perspectives regarding the use of such tools, and educators' perspectives on treating their use as plagiarism, remain unknown. Here, we compare the performance of ChatGPT against students on 32 university-level courses. We also assess the degree to which its use can be detected by two classifiers designed specifically for this purpose. Additionally, we conduct a survey across five countries, as well as a more in-depth survey at the authors' institution, to discern students' and educators' perceptions of ChatGPT's use. We find that ChatGPT's performance is comparable, if not superior, to that of students in many courses. Moreover, current AI-text classifiers cannot reliably detect ChatGPT’s use in school work, due to their propensity to classify human-written answers as AI-generated, as well as the ease with which AI-generated text can be edited to evade detection. Finally, we find an emerging consensus among students to use the tool, and among educators to treat this as plagiarism. Our findings offer insights that could guide policy discussions addressing the integration of AI into educational frameworks.

\section*{Significance statement}
Given the recent emergence of conversational artificial intelligence tools, educational institutions worldwide are facing the significant challenge of addressing the integration of artificial intelligence into educational frameworks. Yet, the literature lacks a systematic study evaluating the performance of such tools on university-level courses and their susceptibility to detection, and also lacks an examination of students' and educators' perspectives on the use of such tools in educational contexts. This work fills these gaps, providing timely and vital insights into the performance of the latest such tool---ChatGPT---and the threat of ``AI-plagiarism'' that it entails. Our findings can inform policy discussions of how to shape student evaluation frameworks in the age of artificial intelligence.

\section*{Introduction}

Generative artificial intelligence (AI) refers to the use of machine learning algorithms that build on existing material, such as text, audio, or images to create new content. 
Recent advancements in this field, coupled with its unprecedented accessibility, has led many to consider it a ``game-changer that society and industry need to be ready for''~\cite{larsen}. In the realm of art, for example, Stable Diffusion and DALL-E have received significant attention for their ability to generate artwork in different styles~\cite{dalle, stablediffusion}. Amper Music, another generative AI tool, is capable of generating music tracks of any given genre, and has already been used to create entire albums~\cite{ampermusic, plaugic_2017}. ChatGPT is the latest tool in this field, which can generate human-like textual responses to a wide range of prompts across many languages. More specifically, it does so in a conversational manner, giving users the ability to naturally build on previous prompts in the form of an ongoing dialogue. This tool has been described as an ``extraordinary hit''~\cite{extraordinary}, and a ``revolution in productivity''~\cite{revolution}, for its seemingly endless utility in numerous out-of-the-box applications, such as creative writing, marketing, customer service, and journalism, just to name a few. The capabilities of the tool have sparked widespread interest, with ChatGPT reaching one million users in only five days after its release~\cite{onemillion}, and soaring to over 100 million monthly users in just two months. 

In spite of its impressive ability, generative AI has been marred by ethical controversies. 
In particular, as generative AI models are trained on massive amounts of data available on the internet, there has been an ongoing debate regarding the ownership of this data~\cite{ai_copyright_broad, lawsuit_1, lawsuit_2}. Furthermore, as these tools continue to evolve, so does the challenge of identifying what is created by humans and what is created by an algorithm. In the context of education, ChatGPT's ability to write essays and generate solutions to assignments has sparked intense discussion concerning academic integrity violations by school and university students. For instance, in the United States its use has been banned by school districts in New York City, Los Angeles, and Baltimore~\cite{shen-berro_2023}. Similarly, Australian universities have announced their intention to return to ``pen and paper'' exams to combat students using the tool for writing essays~\cite{cassidy_2023}. Indeed, many educators have voiced concerns regarding plagiarism, with professors from George Washington University, Rutgers University, and Appalachian State University opting to phase out take-home, open-book assignments entirely~\cite{huang_2023}. In the realm of academic publishing, a number of conferences and journals have also banned the use of ChatGPT to produce academic writing~\cite{vincent_2023, nature_banned}, which is unsurprising given that abstracts written by ChatGPT have been shown to be indistinguishable from human-generated work~\cite{else2023abstracts}. Yet, many have argued for the potential benefits of ChatGPT as a tool for improving writing output, with some even advocating for a push to reform the pedagogical underpinnings of the evaluative process in the education system~\cite{lipman_distler, roose_2023}. Despite these examples, the perspectives of students and educators around the globe remain unclear. The literature also lacks a systematic study comparing the performance of ChatGPT against that of students across various disciplines. Finally, the detectability of ChatGPT-generated solutions to homework remains unknown.

Here, we examine the potential of ChatGPT as a tool for plagiarism by comparing its performance to that of students across 32 university-level courses from eight disciplines. Further, we evaluate existing algorithms designed specifically to detect ChatGPT-generated text, and assess the effectiveness of an obfuscation attack that can be used to evade such algorithms. To better understand the perspectives of students and educators on both the utility of ChatGPT as well as the ethical and normative concerns that arise with its use, we survey participants (N=1601), recruited from five countries, namely Brazil, India, Japan, United Kingdom, and the United States. Additionally, we survey 151 undergraduate students and 60 professors more extensively at the authors' institution to explore differences in perceptions of ChatGPT across disciplines. We find that ChatGPT's performance is comparable, or even superior, to that of students on nine out of the 32 courses. Further, we find that current detection algorithms tend to misclassify human answers as AI-generated, and misclassify ChatGPT answers as human-generated. Worse still, an obfuscation attack renders these algorithms futile, failing to detect 95\% of ChatGPT answers. Finally, there seems to be a consensus among students regarding their intention to use ChatGPT in their schoolwork, and among educators with regard to treating its use as plagiarism. The inherent conflict between these two poses pressing challenges for educational institutions to craft appropriate academic integrity policies related to generative AI broadly, and ChatGPT specifically. Our findings offer timely insights that could guide policy discussions surrounding educational reform in the age of generative AI.

\section*{Results}

We start off by exploring the current capabilities of generative AI to solve university-level exams and homework. To this end, we reached out to faculty members at New York University Abu Dhabi (NYUAD), asking them to provide 10 questions from a course that they have taught at the university, along with three randomly-chosen student answers to each question. Additionally, for each course, ChatGPT was used to generate three distinct answers to each of the 10 questions. Both students' and ChatGPT's answers were then compiled into a single document in random order, labelled as ``Submission~1'' to ``Submission~6''. Each of these submissions were then graded by three different graders, recruited by the faculty member who had taught that course; see Methods for more details, and Supplementary Table~1 for inter-rater reliability. The results of this evaluation can be found in Figure~\ref{fig:grades}a, and see Supplementary Table~2 for the numeric values. Apart from Mathematics and Economics, each discipline has at least one course on which ChatGPT's performance is comparable to, or even surpasses, that of students. These courses are: (i) Data Structures; (ii) Introduction to Public Policy; (iii) Quantitative Synthetic Biology; (iv) Cyberwarfare; (v) Object Oriented Programming; (vi) Structure and Properties of Civil Engineering Materials; (vii) Biopsychology; (viii) Climate/Change; and (ix) Management and Organizations. As a robustness check, we standardized the grades given by each grader of each course to account for grader-specific effects, and again found ChatGPT's performance is comparable, or superior, to student in the above nine courses (Supplementary Table~3).

Having analyzed ChatGPT's performance on different courses, we now perform an exploratory analysis of how its performance varies along different levels of cognition and knowledge. To this end, we asked participating faculty to specify where each of their questions fell along the ``knowledge'' and ``cognitive process'' dimensions of Anderson and Krathwohl's taxonomy~\cite{krathwohl2002revision}; see Table~\ref{tab:taxonomy} for a description of the levels that constitute each dimension. The results of this analysis are summarized in Figure~\ref{fig:grades}b. Interestingly, the gap in performance between ChatGPT and students is substantially smaller on questions requiring high levels of knowledge and cognitive process, compared to those requiring intermediate levels. Also interesting is ChatGPT's performance on questions that require creativity---the highest level along the cognitive process dimension---receiving an average grade of 7.5 compared to the students' average grade of 7.9. Perhaps unsurprisingly, the only questions on which ChatGPT outperforms students are those requiring factual knowledge, attesting to the massive amounts of data on which it was trained. Finally, we compare ChatGPT's performance against different types of questions. To this end, for each question, we asked the participating faculty to specify whether the question: (i) involves mathematics; (ii) involves code snippets; (iii) requires knowledge of a specific author, paper/book, or a particular technique/method; and (iv) is a trick question. The results are summarized in Figure~\ref{fig:grades}c. Again, we find that the largest performance gap between ChatGPT and students was for math-related questions, followed by trick questions. For the time being, humans seem to outperform ChatGPT in these areas.

\begin{table}[htbp]
{\fontsize{11.5}{11.5}\selectfont{
\caption{\textbf{A description of each level along the ``knowledge'' and ``cognitive process'' dimensions of Anderson and Krathwohl's taxonomy, taken from \cite{krathwohl2002revision}.}}
\label{tab:taxonomy}
\begin{center}
\begin{tabular}{ l l }
\toprule
Knowledge dimension &  Description \\
\midrule
Factual & The basic elements that students must know to be acquainted with a discipline\\
 & or solve problems in it \vspace*{0.2cm}\\
Conceptual & The interrelationships among the basic elements within a larger structure that\\
 & enable them to function together\vspace*{0.2cm}\\
Procedural & How to do something; methods of inquiry, and criteria for using skills,\\
 & algorithms, techniques, and methods\vspace*{0.2cm}\\
Metacognitive & Knowledge of cognition in general as well as awareness and knowledge of one’s\\
& own cognition\vspace*{0.2cm}\\
\midrule
Cognitive process dimension &  Description \\
\midrule
Remember & Retrieving relevant knowledge from long-term memory\vspace*{0.2cm}\\
Understand & Determining the meaning of instructional messages, including oral, written,\\
 & and graphic communication\vspace*{0.2cm}\\
Apply & Carrying out or using a procedure in a given situation\vspace*{0.2cm}\\
Analyze & Breaking material into its constituent parts and detecting how the parts relate\\
 & to one another and to an overall structure or purpose \vspace*{0.2cm}\\
Evaluate & Making judgments based on criteria and standards\vspace*{0.2cm}\\
Create & Putting elements together to form a novel, coherent whole or make\\
& an original product\vspace*{0.2cm}\\
\bottomrule
\end{tabular}
\end{center}
}}
\end{table}

\begin{figure}[htpb]
    \centering
    \includegraphics[width=\linewidth]{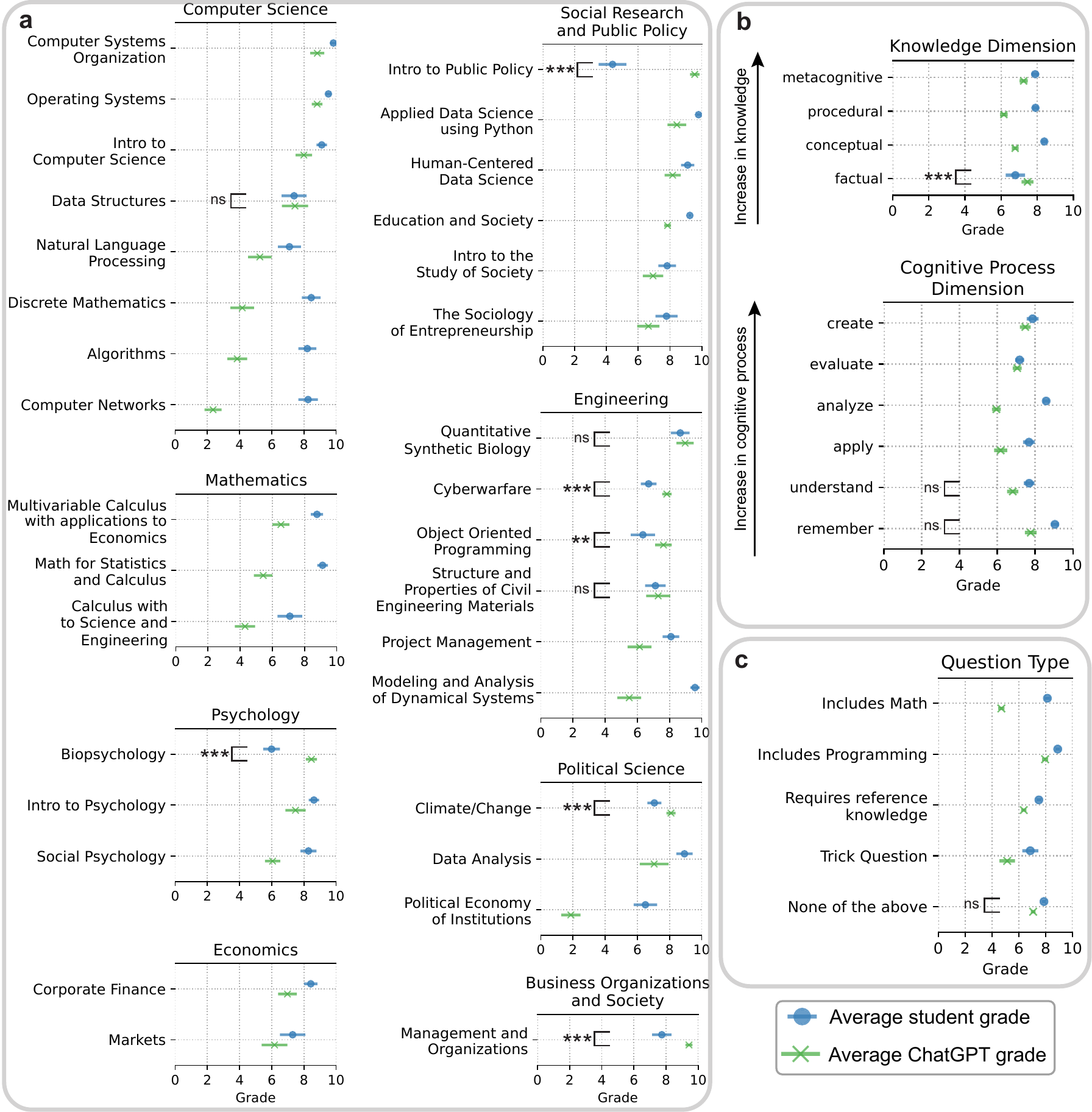}
    \caption{\textbf{Comparing ChatGPT to university-level students.} Comparing ChatGPT's average grade (green) to the students' average grade (blue), with error bars representing 95\% confidence intervals. (\textbf{a}) Comparison across university courses. (\textbf{b}) Comparison across the ``cognitive process'' and ``knowledge'' dimensions of Anderson and Krathwohl taxonomy's of learning. (\textbf{c}) Comparison across question types. $p$-values are calculated using bootstrapped two-sided Welch's T-test, and only shown for courses where GPT does not receive a significantly lower grade compared to students (** = $p < 0.01$; *** = $p < 0.001$; $ns = $ not significant, i.e., $p > 0.05$).}
    \label{fig:grades}
\end{figure}

\begin{figure}[htpb]
    \centering
    \includegraphics[width=\linewidth]{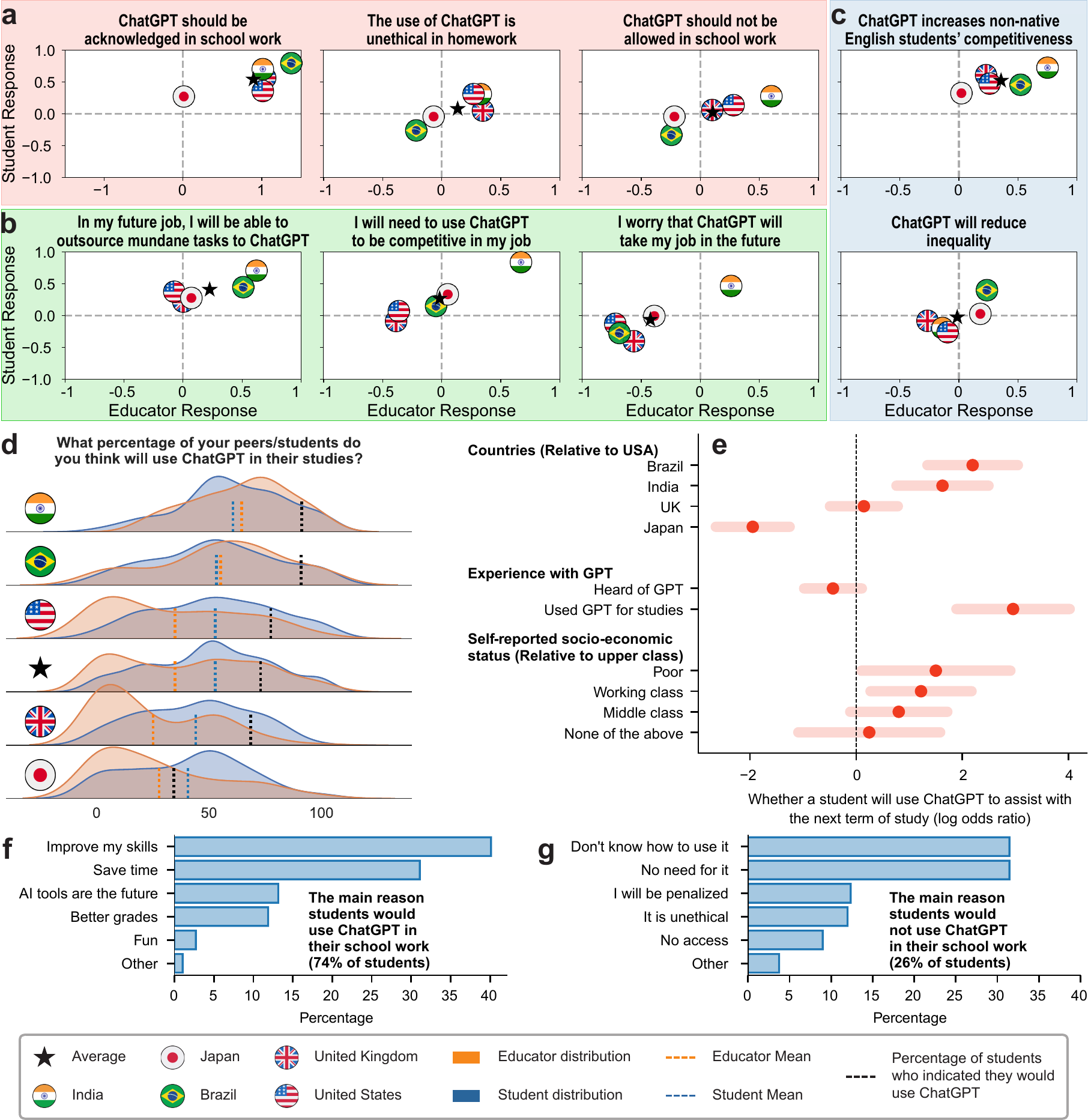}
    \caption{\textbf{Global survey responses.} (\textbf{a}-\textbf{c}) Educators' average responses (x-axis) and students' average responses (y-axis) to eight questions regarding ChatGPT;  $1 =$ agree; $-1 =$ disagree; $\star =$ average over the five countries. (\textbf{d}) Distributions of students'/educators' estimation of the percentage of their peers/students who they believe will use ChatGPT in their studies. The dashed black line represents the percentage of students who indicated they would use ChatGPT in their studies. (\textbf{e}) OLS-estimated coefficients and 95\% confidence intervals of selected independent variables predicting a student's decision to use ChatGPT in their next term of studies; see Supplementary Materials for all independent variables considered. (\textbf{f}-\textbf{g}) A breakdown of why students indicated they would or would not use ChatGPT.}
    \label{fig:survey}
\end{figure}

To understand how the use of ChatGPT is perceived by educators and students, we fielded a global survey in five countries, namely Brazil, India, Japan, the UK, and the US, targeting a minimum of 100 educators and 200 students per country; see Methods for more details. A summary of our findings can be seen in Figure~\ref{fig:survey}. Before delving into this analysis, it should be noted that students and educators in our survey come from various levels of the education. As such, we performed a similar analysis focusing only on undergraduate and postgraduate students, as well as university-level educators, and found broadly similar results; see Supplementary Figure~1. We start off by comparing the responses of students vs.\ educators in different countries; see panels a-c in Figure~\ref{fig:survey}. Here, each plot corresponds to a different question in the survey, asking about the degree to which respondents agree or disagree with a particular statement about ChatGPT ($-2 =$ strongly disagree; $-1 =$ disagree; $0 =$ neutral; $1 =$ agree; $2 =$ strongly agree). We present statements in three broad categories: (i) the ethics of using ChatGPT in the context of education in panel~a; (ii) the impact of ChatGPT on future jobs in panel~b; and (iii) the impact of ChatGPT on inequality in education in panel~c; see Supplementary Figure~4 for the remaining statements on the survey. Starting with panel~a (ethics), there seems to be a consensus that using ChatGPT in school work should be acknowledged. In contrast, opinions vary when it comes to determining whether the use of ChatGPT in homework is unethical and whether it should be prohibited in school work, e.g., students in India and the US believe it is unethical and should be prohibited, while those in Brazil believe otherwise. Moving on to panel~b (jobs), students in all five countries believe they can outsource mundane tasks to ChatGPT, and the educators in Brazil and India seem to agree with this statement. India is the only country where educators believe ChatGPT is needed to increase their competitiveness in their job, and the students therein agree the most with this statement. Moreover, educators and students in India are the only ones who worry that ChatGPT will take their job in the future. In terms of panel~c (inequality), there seems to be a consensus that ChatGPT will increase the competitiveness of non-native English students. When it comes to whether ChatGPT will reduce inequality in education, educators in Brazil and Japan (the two non-English speaking countries in our sample) agree with this statement, while those in the remaining three countries disagree.

Next, we compare the distribution of the educators' and students' responses to the following question: ``What percentage of your students/peers do you think will use ChatGPT in their studies?''. The results are depicted in Figure~\ref{fig:survey}d, where the distributions of educators and students responses are illustrated in orange and blue, respectively, with vertical lines of the same colors representing the means. The black vertical line represents the percentage of students who answered ``Yes'' to the question: ``Considering your next term of studies, would you use ChatGPT to assist with your studies?''. As can be seen in the fourth row, representing the average response across the five countries, 74\% of students indicate that they would use ChatGPT (black line), while both educators and students underestimated this share. For the students who said they would use it (74\%), their main reasons are to improve their skills and save time (Figure~\ref{fig:survey}f). As for those who said they would not use ChatGPT (26\%), their main reasons are not knowing how to do so or not having a need for it, rather than a fear of being penalized or acting unethically (Figure~\ref{fig:survey}g). 

Finally, we perform an OLS regression analysis to explore which factors that might be associated with the students' decision to use ChatGPT during their next term of studies. Figure~\ref{fig:survey}e summarizes the results for a few independent variables of interest; the remaining variables can be found in Supplementary Figure~2-3. As can be seen, students from Brazil and India are significantly more likely, and students from Japan are significantly less likely, to use ChatGPT than those from the USA. As for prior experience with ChatGPT, those who have used it are significantly more likely to use it again. On the contrary, simply hearing about ChatGPT is not significantly associated with the students' decision to use it to assist with their studies. Finally, students from poor and working class backgrounds are significantly more likely to indicate they would use ChatGPT for their studies compared to upper class students.

\begin{figure}[htpb]
    \centering
    \includegraphics[width=\linewidth]{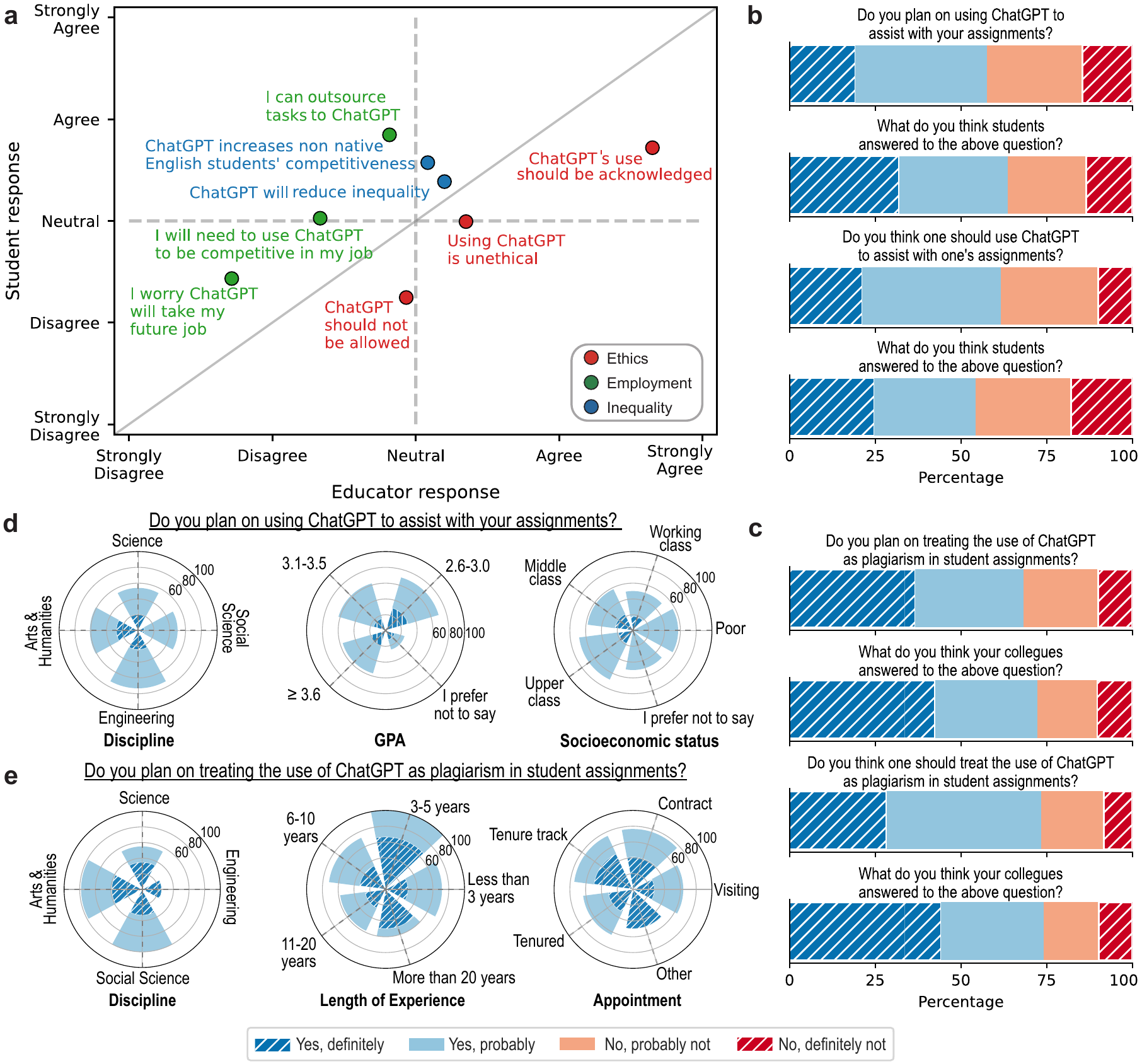}
    \caption{\textbf{NYUAD survey responses.}  (\textbf{a}) Professors' responses (x-axis) and Students' responses (y-axis) to eight statements regarding ChatGPT; the statements about ethics, employment, and inequality are colored in red, green, and blue, respectively. (\textbf{b} - \textbf{c}) Descriptive norms and Injunctive norms of students and professors at NYUAD with regards to using ChatGPT, and treating it as plagiarism, respectively. (\textbf{d}) Comparing student responses across disciplines, GPA, and socioeconomic status. (\textbf{e}) Comparing professor responses across disciplines, length of teaching experience, and type of appointments.}
    \label{fig:nyuadsurvey}
\end{figure}

Having analyzed the global survey, we now shift our attention to the second survey, which was conducted at the authors' institution---NYUAD. While this survey is narrower in scope than the previous one, it focuses on university students and professors, allowing us to examine variances in responses with respect to GPA in the case of students, and type of appointment in the case of professors. Figure~\ref{fig:nyuadsurvey}a depicts the responses of 151 students (y-axis) and 60 professors (x-axis) with regards to the same eight statements discussed earlier, grouped into three broad categories: (i) the ethics of using ChatGPT in the context of education in red; (ii) the impact of ChatGPT on future jobs in green; and (iii) the impact of ChatGPT on inequality in education in blue; see Supplementary Figure~5 for the remaining statements. As can be seen, professors agreed more, or disagreed less, with all statements in category (i) compared to students (notice how all red data points fall below the diagonal) while students agreed more, or disagreed less, with all statements in categories (ii) and (iii). Despite these differences, professors and students seem to agree that ChatGPT's use should be acknowledged, and neither believes that it will take their future job.

Figure~\ref{fig:nyuadsurvey}b shows students' empirical expectations (first and third rows) and normative expectations (second and fourth rows) of whether they plan to use ChatGPT to assist with their assignments (first two rows) and whether they think one should use ChatGPT to assist with one's assignment (last two rows). The majority of students plan to use ChatGPT to assist with their assignments (57\%), and expect their peers to use it for this purpose (64\%). Moreover, the majority believe ChatGPT should be used (61\%), and expect that their peers believe it should be used (55\%), to assist with assignments. Similarly, Figure~\ref{fig:nyuadsurvey}c depicts professors' empirical expectations (first and third rows) and normative expectations (second and fourth row) of whether they plan to treat ChatGPT's use as plagiarism (first two rows) and whether they think one should treat ChatGPT's use as plagiarism (last two rows). Most professors plan to treat the use of ChatGPT as plagiarism (69\%), and expect others to do so (71\%). Furthermore, the majority believe the use of ChatGPT should be treated as plagiarism (72\%), and expect that their peers believe it should be treated as such (73\%).

Figure~\ref{fig:nyuadsurvey}d compares students' intention to use ChatGPT for their studies across disciplines, GPA, and socioeconomic status. Starting with discipline, the majority of students from all four disciplines indicated that they plan on using ChatGPT. As for GPA, disregarding those who preferred not to disclose their GPA, the majority of students from all GPA brackets into which students fell indicated that they would use the tool.
Similarly, with regards to socioeconomic status, the majority of students from all socioeconomic mentioned they would use ChatGPT. As for professors, Figure~\ref{fig:nyuadsurvey}e compares responses on whether they plan to treat the usage of ChatGPT as plagiarism across disciplines, length of teaching experience, and type of appointment at the institution. As shown in this figure, in each discipline apart from Engineering, the majority of faculty intend to treat it as plagiarism. In terms of teaching experience, again, the majority of faculty intend to do so, regardless of their experience. Similarly, for each type of appointment at the institution, the majority of faculty plan on treating the use of ChatGPT as plagiarism.

\begin{figure}[htpb]
    \centering
    \includegraphics[width=\linewidth]{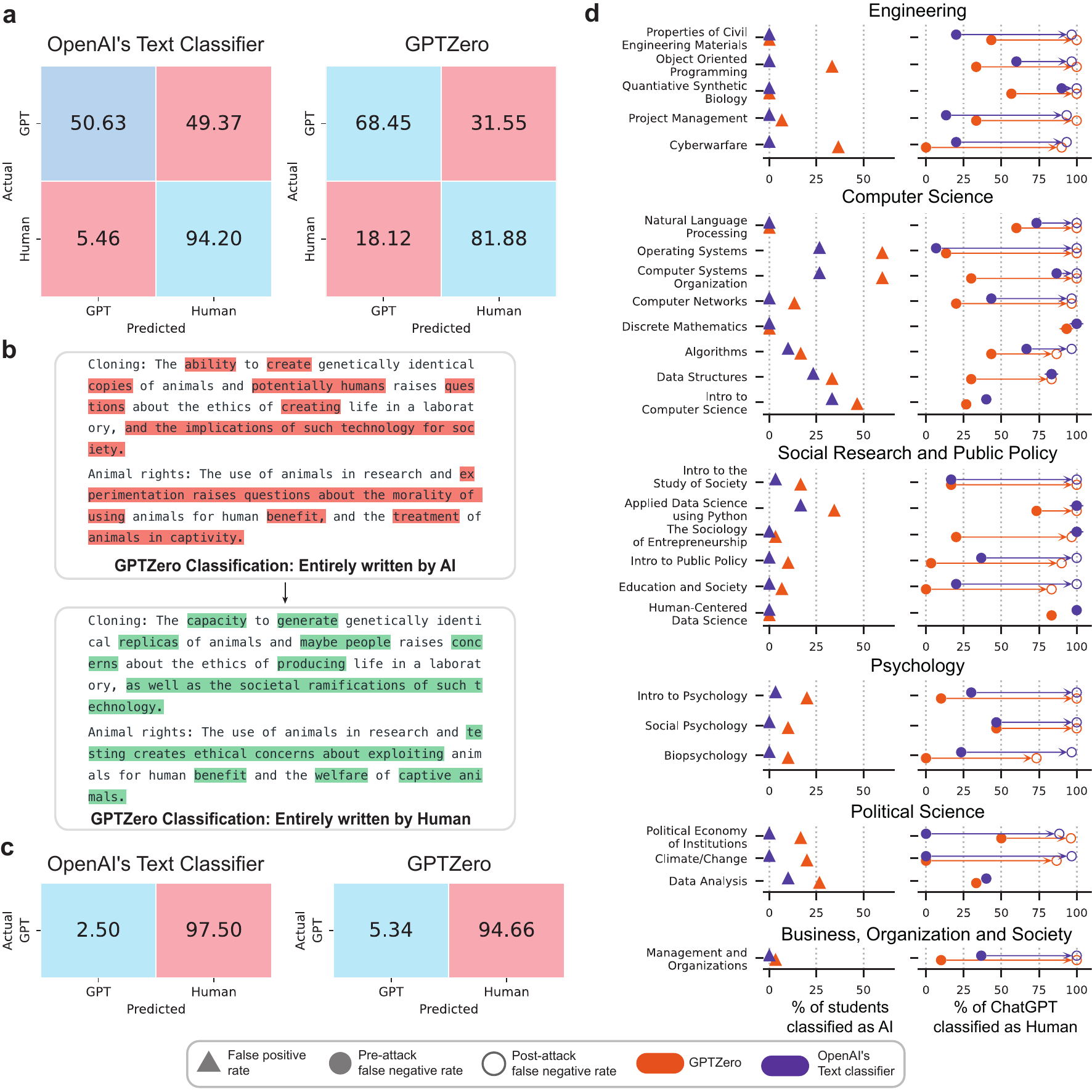}
    \caption{\textbf{Evaluating AI-text detectors.} Quantifying the degree to which \textit{GPTZero} and \textit{AI text classifier} can identify whether a body of text is AI-generated. (\textbf{a}) Normalized confusion matrices of \textit{GPTZero} and \textit{AI text classifier}. (\textbf{b}) An example of the Quillbot attack taken from our dataset, with changes highlighted in different colors. (\textbf{c}) Normalized confusion matrices after deploying the Quillbot attack. (\textbf{d}) Evaluating the two classifiers on different courses.}
    \label{fig:obfuscation}
\end{figure}

We conclude our analysis by assessing the detectability of using ChatGPT. To this end, we use two classifiers, namely \textit{GPTZero}~\cite{GPTzero} and OpenAI's own \textit{AI text classifier}~\cite{openaiclassifier}, both of which are designed specifically to determine whether a body of text has been generated using AI. We use these classifiers to quantify the percentage of human submissions that are misclassified as ChatGPT, as well as the proportion of ChatGPT submissions that are misclassified as human. More formally, our goal is to construct a normalized confusion matrix for each classifier, and to quantify its false-positive and false-negative rates; see Methods for more details. Out of all 320 questions in our dataset, there were 40 questions whose answers consisted entirely of mathematical equations that are incompatible with the classifiers; those were excluded from our evaluation, leaving us with 280 questions. The results are depicted in Figure~\ref{fig:obfuscation}a. As can be seen, OpenAI's Text Classifier misclassifies 5\% of student submissions as AI-generated, and 49\% of ChatGPT's submissions as human-generated. GPTZero has a higher false positive rate (18\%), but a lower false-negative rate (32\%).

Next, we analyze the degree to which these classifiers are susceptible to obfuscation attacks. We devised a simple attack that involves running the ChatGPT-generated text through Quillbot \cite{quillbot}---a popular paraphrasing/re-writing tools utilized by students worldwide. Our ``Quillbot attack'' was chosen since it can be readily implemented by students, without requiring any technical know-how; see Methods for a detailed description of the attack, and see Figure~\ref{fig:obfuscation}b for a stylized example of how it works. Out of the 280 questions, 47 required answers that involved snippets of computer code. For those questions, about half of the ChatGPT submissions were already misclassified as human-generated, without any obfuscation attacks. Based on this, as well as the fact that Quillbot does not work on code snippets, we excluded those questions from the sample on which we run the Quillbot attack. The results of this evaluation are summarized in Figure~\ref{fig:obfuscation}c. As shown in this figure, the attack is remarkably effective. In the case of OpenAI's text classifier, the false-negative rate increased from 49\% to 98\%, and in the case of GTPZero, the false-negative rate increased from 32\% to 95\%. Figure~\ref{fig:obfuscation}d evaluates the two classifiers on various courses. In the left column, triangles represent the percentage of student submissions that are misclassified as AI-generated. In the right column, the solid and empty circles represent the percentage of ChatGPT submissions that are misclassified as human-generated before, and after, the Quillbot attack, respectively, with arrows highlighting the difference between the two circles. Note that there are a handful of courses for which there are no empty circles since they include code snippets. As the figure illustrates, in each discipline, there are courses for which students' submissions are misclassified, as well as courses for which ChatGPT submissions are misclassified. Importantly, the Quillbot attack is successful on all courses, often increasing the false-negative rate to 100\%. Supplementary Figure~6 analyzes the performance of both classifiers on different question types, and analyzes their performance along both dimensions of the Anderson and Krathwohl taxonomy~\cite{wilson2016anderson}. 

\section*{Discussion}
This study aimed to provide both quantitative and qualitative insights in the perceptions and performance of current text-based generative generate AI in the context of education. We surveyed educators and students at the authors' institution and in five countries, namely Brazil, India, Japan, UK, and USA, regarding their perspectives on ChatGPT. There is a general consensus between educators and students that the use of ChatGPT in school work should be acknowledged, and that it will increase the competitiveness of students who are non-native English speakers. Additionally, there is a consensus among students that, in their future job, they will be able to outsource mundane tasks to ChatGPT, allowing them to focus on substantive and creative work. Interestingly, the majority of surveyed students expressed their intention to use ChatGPT despite ethical considerations. For example, students in India think the use of ChatGPT in homework is unethical and should not be allowed, while those in Brazil think it is ethical and should be allowed, yet in both countries, the vast majority of surveyed students (94\%) indicated that they would use it to assist with their homework and assignments in the coming semester. The negative implications of this finding go beyond the university, as studies have demonstrated that academic misconduct in school is highly correlated to dishonesty and with similar decision-making patterns in the workplace~\cite{harding2003examination, nonis2001examination}.

The NYUAD survey allowed us to examine whether there exists a descriptive norm among students backed by injunctive  norms regarding the use of ChatGPT, and among professors with regards to treating ChatGPT's use as plagiarism. More specifically, we investigate whether individuals prefer to conform to these behaviours on the condition that they expect the majority of their peers would also conform to them~\cite{cialdini1990focus,bicchieri2010norms, bicchieri2005grammar}. Our results indicate that the majority of students plan to use ChatGPT to assist with their assignments, and believe their peers would approve of its use, implying that the use of ChatGPT is likely to emerge as a norm among students. Similarly, we find that the majority of professors plan to treat the use of ChatGPT as plagiarism tool, and believe their peers would approve of treating it as such, implying that the treatment of ChatGPT as plagiarism is likely to emerge as a norm among professors. The inherent conflict between these two could create an environment where students' incentives suggest they should hide the use of ChatGPT while professors are eager to detect its use (and may expect their colleagues to do the same). This poses a major challenge for educational institutions to craft appropriate academic integrity policies related to generative AI broadly, and ChatGPT specifically.

When comparing ChatGPT's performance to that of students across university-level courses, the results indicate a clear need to take ``AI-plagiarism'' seriously. Specifically, ChatGPT's performance was comparable, if not superior, to students' performance on 12 (38\%) out of the 32 courses considered in our analysis. These findings, coupled with the clear ease with which students can avoid detection by current AI-text classifiers as demonstrated in our study, suggests that educational institutions are facing a serious threat with regards to current student evaluation frameworks. To date, it is unclear whether to integrate ChatGPT into academic and educational policies, and if so, how to enforce its usage. While several school districts, and even an entire country in the case of China, have already banned access to the tool~\cite{schwartz_2022}, the degree to which these measures can be effectively enforced remains unknown. Treating the use of ChatGPT as plagiarism is likely to be the most common approach to counter its usage. However, as we have shown, current measures to detect its usage are futile. Other universities have opted to change their evaluative processes to combat the tool's usage, by returning to in-person hand-written and oral assessments~\cite{cassidy_2023}---a step that may undo decades of advancements aimed at deploying alternatives to high stakes examinations that disadvantage some categories of students and minorities~\cite{johnson2008stop}. Our findings suggest that ChatGPT could disrupt higher education's ability to verify the knowledge and understanding of students, and could potentially decrease the effectiveness of teacher feedback to students. In education, it is crucial to confirm that what is being assessed has been completed by the person who will receive the associated grades. These grades go on to influence access to scholarships, employment opportunities, and even salary ranges in some countries and industries~\cite{kittelsen2017grades}. Teachers use assessment to give feedback to learners, and higher education institutions use the assessment process to credential learners. The value of the degree granted is in part linked to the validity of the assessment process. Countermeasures to academic misconduct such as plagiarism include applying detection software, such as Turnitin, and emphasising honor codes in policy and cultural norms. Higher education, in particular, is susceptible to disruption by ChatGPT because it makes it very difficult to verify if students actually have the skills and competencies their grades reflect. It disrupts how learning is assessed because demonstrated learning outcomes may be inaccurate and thus, prolific use could diminish the academic reputation of a given higher education institution. Our findings shed light on the perceptions, performance, and detectibility of ChatGPT, which could guide the ongoing discussion on whether and how educational policies could be reformed in the presence of not only ChatGPT, but also its inevitably more sophisticated successors.

\section*{Methods}

\subsection*{ChatGPT performance analysis}
All faculty members at New York University Abu Dhabi (NYUAD) were invited to participate in this study and were asked to provide student assessments from their courses. In particular, they were asked to follow these steps: (i) select a course that they have taught at NYUAD; (ii) randomly select 10 text-based questions from labs, homework, assignments, quizzes, or exams in the course; (iii) randomly select three student submissions for each of the 10 questions from previous iterations of the course; (iv) convert any hand-written student submissions into digital form; (v) obtain students' consent for the usage of their anonymized submissions in the study; and (vi) obtain students' confirmation that they were 18 years of age or older. 

When choosing their questions, participating faculty were given the following guidelines: (i) Questions should not exceed 2 paragraphs (but the answer could be of any length); (ii) No multiple-choice questions were allowed; (iii) No images, sketches, or diagrams were allowed in the question nor the answer; (iv) No attachments are allowed in the question nor the answer (i.e., a paper or a report to be critiqued, or summarized etc.); (v) The question could include a table, and the answer could have a table to be filled; (vi) The question and the answer could include programming code; (vii) The question and the answer could include mathematics (e.g., equations); and (viii) Fill-the-gap questions were allowed.

For each course, ChatGPT was used to generate three distinct answers to each of the 10 questions. This step was carried out between the 13th and 29th of January, 2023. Both students' and ChatGPT's answers were then compiled into a single document in random order, labelled as ``Submission~1'' to ``Submission~6''. The faculty responsible for each course was asked to recruit three graders (either teaching assistants, postdoctoral fellows, research assistants, PhD students, or senior students who have taken the course before and obtained a grade of A or higher), and to provide the graders with rubrics and model solutions to each question. The participating faculty was instructed not to inform the graders that some of the submissions are AI-generated. Each grader was compensated for their time at a rate of \$15 per hour. The participating faculty vetted the graders' output and ensured that the grading is in line with the rubrics.

For each of their 10 questions, the faculty were asked to answer the following: (i) Where does the question fall along the ``knowledge'' dimension and the ``cognitive process'' dimension of the Anderson and Krathwohl taxonomy~\cite{krathwohl2002revision}; (ii) Does the question or the answer involve mathematics? (Yes/No); (iii) Does the question or the answer involve snippets of computer code? (Yes/No); (iv) Does the question or the answer require knowledge of a specific author, scientific paper/book, or a particular technique/method? (Yes/No); and (v) Is the question a trick question? (Yes/No). To explain what a trick question is, the following description was provided to participating faculty: ``A trick question is a question that is designed to be difficult to answer or understand, often with the intention of confusing or misleading the person being asked. Trick questions are often used in games and puzzles, and can be used to test a person's knowledge or problem-solving skills.''

\subsection*{Survey}
Participants were recruited from Brazil, India, Japan, United Kingdom, and United States, with a minimum of 200 students and 100 educators per country; see Supplementary Table~4-5 for descriptive statistics, and Supplementary Note~1.1-1.2 for the complete survey instrument. The survey was piloted in the United States on Prolific with open-ended questions to investigate that respondents did not find anything odd or surprising about survey questions, and also to ensure no new topics emerged about ChatGPT that the opinion statements covered. Data collection was conducted using the SurveyMonkey platform. Studies have shown the validity of using this platform to conduct surveys used in academic research, in addition to the numerous studies which use such platforms for conducting surveys globally~\cite{bentley2017comparing, mielke2017ideals, parsa2014obstacles, evans2009developing, peloquin2010measuring}. The platform utilizes email and location verification for fraud detection and ID exclusions, thereby guarding against duplicate and bot-submitted responses.  SurveyMonkey enables the targeting of samples with pre-specified demographics. Specifically, in the US, we identified educators as those whose job is ``educator (e.g., teacher, lecturer, professor),'' and identified students as those who had a ``student status.'' As for the remaining four countries, the above filters were not available on SurveyMonkey. Therefore, we recruited 100 respondents whose primary role is ``Faculty/Teaching Staff'' and recruited 200 respondents whose occupation was ``studies.'' In addition, we ensured that the analytical sample only contains current students and educators who are either teachers or professors based on respondents’ self-reports. Respondents filled out the survey in the official language of their country. The global surveys were conducted on the 19th and 20th of January, 2023. 

As for the NYUAD survey, it includes responses from 151 students (out of a total of 2103, 7.2\%) and 60 professors (out of a total of 393, 15.2\%) collected between January 23rd and February 3rd, 2023. This survey used the same questions as those used in the global survey, but with a few additions. Specifically, for students, we asked them to specify their current cumulative GPA, year of study, as well as empirical and normative expectations~\cite{bicchieri2005grammar, bicchierri_2016} related to the use of ChatGPT among students. As for faculty, we asked them about the number of years they have been teaching, their academic division, their appointment type (e.g., tenure, tenure track, contract, or visiting), as well as their empirical and normative expectations surrounding the treatment of ChatGPT as plagiarism. We excluded student participants who indicated that they aware of our study, as well as faculty members who coauthored the paper. Descriptive statistics can be found in Supplementary Tables~6-7, while the survey questions can be found in Supplementary Notes~1.3-1.4.

\subsection*{Obfuscation attacks}
Once all student and ChatGPT submissions were collected for all courses, they were used to test two classifiers designed specifically to determine whether a body of text has been generated using AI, namely, \textit{GPTZero}~\cite{GPTzero}, and OpenAI's own \textit{AI text classifier}~\cite{openaiclassifier}. The former classifies text into five categories: (i) ``likely to be written entirely by a human'', (ii) ``most likely human written but there are some sentences with low perplexities'', (iii) ``more information is needed'', (iv)``includes parts written by AI'', (v) ``likely to be written entirely by AI''.
As for the latter, it classifies text into five categories: (i) very unlikely, (ii) unlikely, (iii) unclear if it is, (iv) possibly, or (v) likely AI-generated. We start our analysis by examining the proportion of human submissions that are misclassified as ChatGPT, as well as the proportion of ChatGPT submissions that are misclassified as human. More formally, our goal is to construct a normalized confusion matrix for each classifier, to quantify its false-positive and false-negative rates. For both classifiers, ChatGPT submissions that are classified as categories (i) or (ii) are treated as false negatives, while human submissions that are classified as (iv) or (v) are treated as false positives. 
As for submissions classified as (iii), we consider them to be classified as human. This decision is motivated by our context, where the classifier is used to detect AI-plagiarism. Given that a student is arguably ``innocent until proven otherwise'', having an inconclusive classification means that, as far as the teacher is concerned, there is no conclusive evidence incriminating the student of AI-plagiarism, implying that such a classification has practically the same consequences as being classified as human. Based on this, ChatGPT submissions that are classified as (iii) are treated as false negatives.

To evaluate the classifiers' susceptibility to obfuscation attacks, we devised a simple attack that involves running the ChatGPT-generated text through Quillbot. Our attack starts by setting the ``Synonym'' slider in Quillbot to the highest level, which maximizes the number of modifications introduced to the text. Then, we run every ChatGPT-generated submission through each of the ``Modes'' available on Quillbot, each resulting in a different output. Each output is then re-analyzed on both classifiers, and if any of them manages to flip the text classification from category (iv) or (v) to any other category, the attack on this particular submission is considered successful.

\section*{Data availability}
All of the data used in our analysis can be found at the following repository: \url{https://github.com/comnetsAD/ChatGPT}.

\section*{Author Contributions}
T.R. and Y.Z. conceived the study and designed the research;
H.I., B.B, N.G., T.R., and Y.Z. wrote the manuscript;
H.I., BA.A., and B.B performed the literature review;
H.I., T.R., and Y.Z. collected the data;
K.M. designed the surveys and ran the NYUAD survey;
H.I. and Y.Z. ran the global survey;
H.I., F.L., R.A., S.B., T.R., and Y.Z. analyzed the data and ran the experiments;
H.I., F.L., R.A., S.B., T.R., and Y.Z. designed the obfuscation attack;
H.I., F.L., R.A., S.B., and Y.Z. ran the obfuscation attack;
H.I., F.L., T.R., and Y.Z. produced the visualizations;
W.A., T.A., BE.A., R.B., J.B., E.B., K.C., M.C., M.D., B.G.S., Z.E., D.F., AL.G., AN.G., N.H., A.H., A.K., L.K., K.K., K.L., S.S.L., S.M., M.M., D.M., A.M., M.P., A.R., D.Z., N.G., K.M., T.R., and Y.Z. provided course questions and answers, and recruited graders.

\section*{Ethics statement}
The research was approved by the Institutional Review Board of New York University Abu Dhabi (HRPP-2023-5). All research was performed in accordance with relevant guidelines and regulations. Informed consent was obtained from all participants in every segment of this study.

\section*{Acknowledgement}
K.M. acknowledges funding from the NYUAD Center for Interacting Urban Networks (CITIES), funded by Tamkeen under the NYUAD Research Institute Award CG001.

\bibliography{scibib.bib}
\bibliographystyle{naturemag}

\end{document}